\documentclass[a4paper]{article}

\usepackage[english]{babel}
\usepackage[utf8]{inputenc}
\usepackage{amsmath}
\usepackage{graphicx}
\usepackage[colorinlistoftodos]{todonotes}
\usepackage{caption}
\usepackage{subcaption}
\usepackage{soul}

\title{A Retrospective Recount of Computer Architecture Research with a Data-Driven Study of Over Four Decades of ISCA Publications}

\author{Omer Anjum \and Wen-Mei Hwu \and Jinjun Xiong}
\date{%
    IBM-ILLINOIS Center for Cognitive Computing Systems Research\\ [2ex]
    \today
}
\begin{document}
\maketitle

The ACM/IEEE International Symposium on Computer Architecture (ISCA) conference is one of the premier forums for presenting, debating and advancing new ideas and experimental results in computer architecture. According to recent calls for papers, research topics are solicited on a broad range of areas, including, but not limited to: processors, memories, storage systems, architectures, interconnection networks, instructions, thread-level parallelism, data-level parallelism, dependable architectures, architecture support for parallel software development, architecture support for security, power and energy efficient architectures, application specific architectures, reconfigurable architectures, embedded architectures, network and router architectures, architectures for emerging technologies, architecture modeling and performance evaluation.
\begin{figure}[h!]
  \centering
    \includegraphics[width=\linewidth]{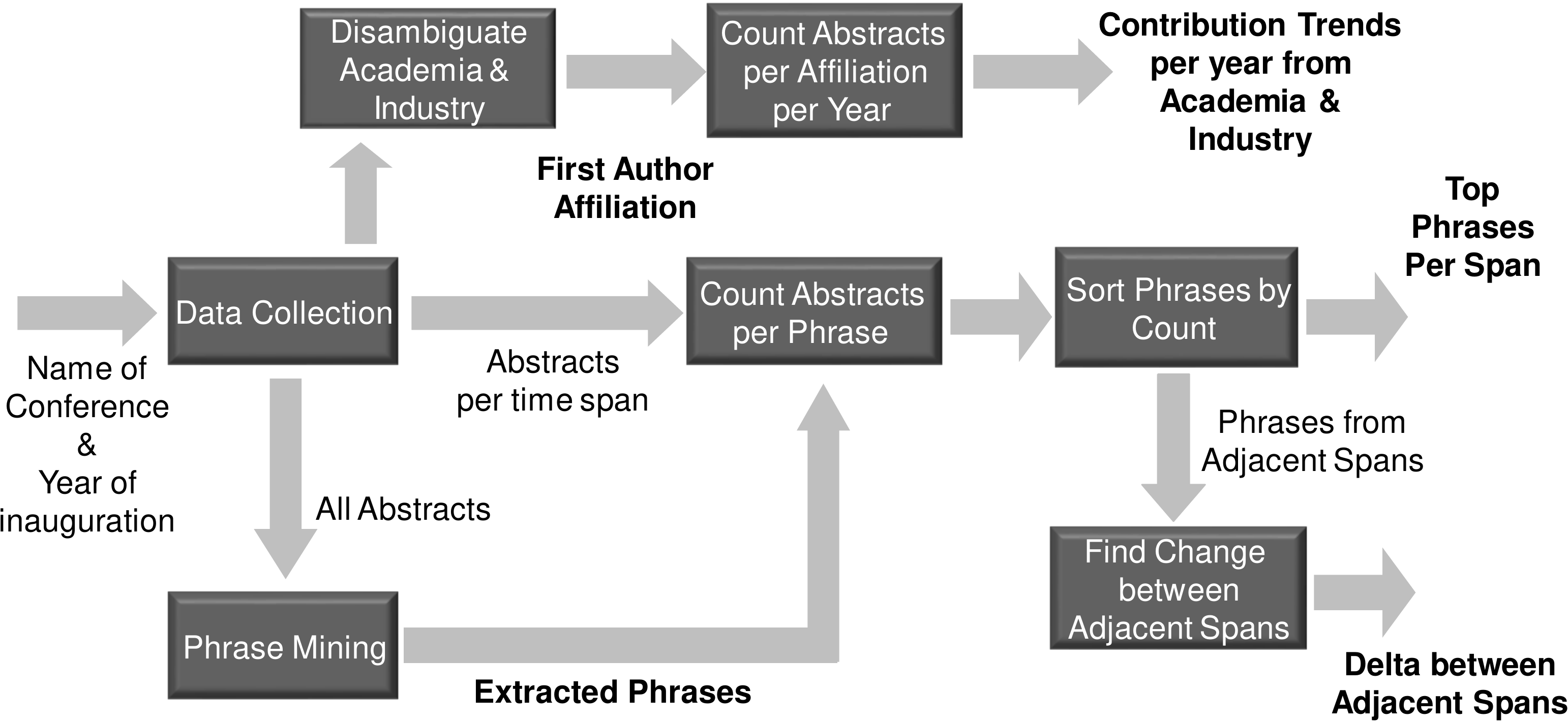}
    \caption{An Overview of the Pipeline}
  \label{fig:pipeline}
\end{figure}

Recently we conducted a study of some notable publication trends for ISCA from 1973, when it was inaugurated, to 2018. The main questions we were trying to answer was how the topics and thus the community interests evolved over these 45 years. Our data set includes all the abstracts of papers published in the conference. The source of our data set is Microsoft Academic Graph \cite{MAG}.  The pipeline we developed for producing the results is shown in Figure \ref{fig:pipeline}. We will discuss key features and current limitations of the pipeline in a later section of this blog. We plan to make the pipeline available as an open source project to enable similar studies for other conferences and domains.

\section{Longitudinal Observations}
\begin{figure}[h!]
  \centering
    \includegraphics[width=0.9\linewidth]{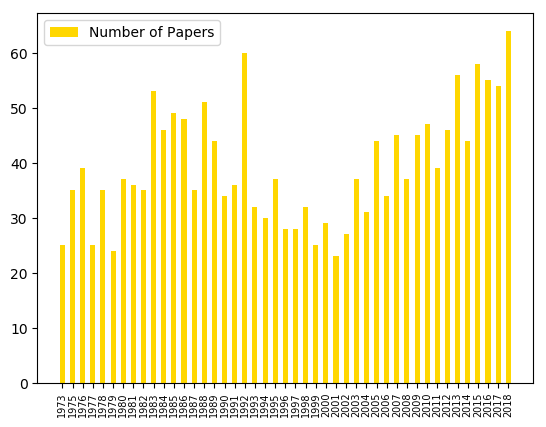}
    \caption{Number of Papers Each Year}
  \label{fig:iscaNumberOfSubmissions}
\end{figure}

\begin{figure}[h!]
  \centering
    \includegraphics[width=0.9\linewidth]{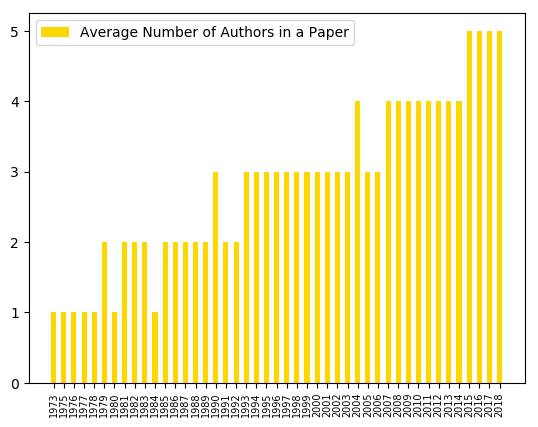}
    \caption{Average Number of Authors in a Paper}
  \label{fig:iscaAvgNumAuth}
\end{figure}

\begin{figure}[h!]
  \centering
    \includegraphics[width=\linewidth]{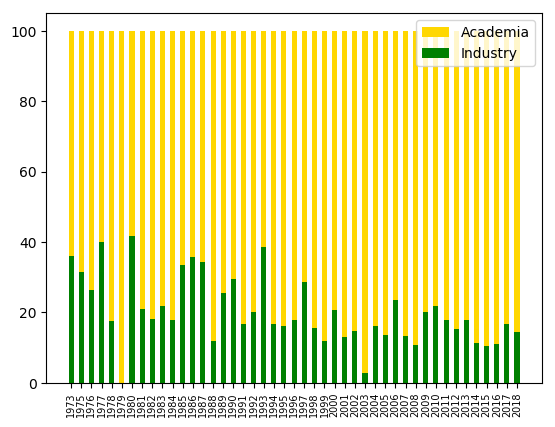}
    \caption{Percentage of industry vs. academia affiliation of first authors}
  \label{fig:AcademiaVsIndustry}
\end{figure}

We first show the number of papers published at ISCA 
in Figure~\ref{fig:iscaNumberOfSubmissions}, where we observe 
an increase in the number of papers published in 1980s and then in 2010s. Highest peak is observed in 1992 and 2018. It would be interesting to see if the number of papers published at ISCA would decline in the coming years completing another cycle. In Figure~\ref{fig:iscaAvgNumAuth} we observe that the average number of authors per paper has also increased over time. According to the trends in the past it appears that the existing trend may continue for few more years and the average number of authors per submission will likely increase. %In Figure~\ref{fig:topichistory} we show the trends of some of the key topics to give a summarized view of activities happening in the ISCA community. 
In Figure~\ref{fig:AcademiaVsIndustry}, we see that the there is a long-term trend in decline of papers by first authors from industry. We believe that this trend is caused by two important forces. First, there is a decline of publication-oriented research activities in the industry in favor of product development. Second, the amount of ``paper engineering'' effort to write a competitive paper for acceptance by ISCA has been increasing over time. While graduate students can spend the effort, it is not clear if industry researchers can justify such efforts.

\begin{figure}[h!]
  \centering
    \includegraphics[width=\linewidth]{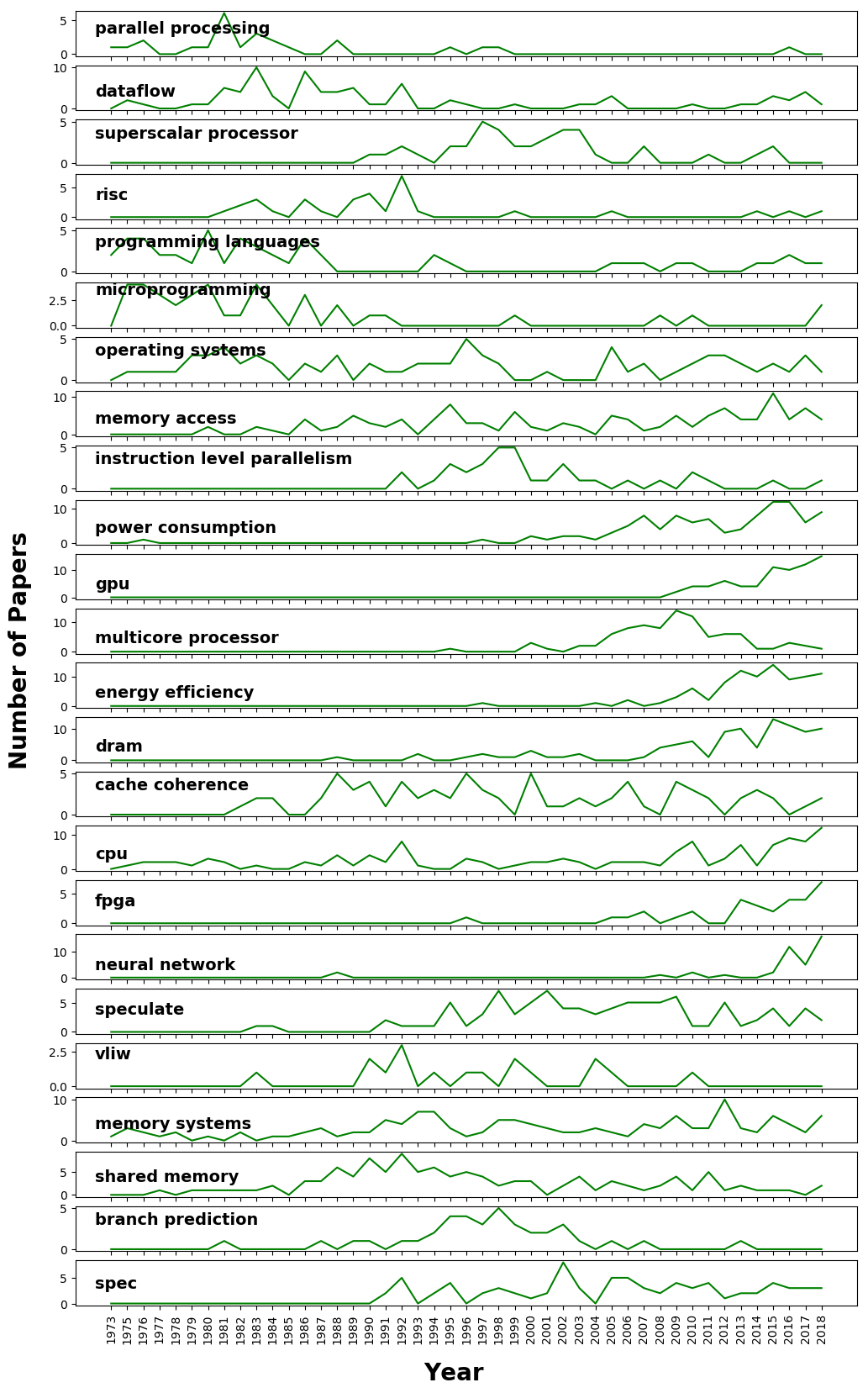}
    \caption{A long-term history of selected top topics}
  \label{fig:topichistory}
\end{figure}

We then make a few observations on longitudinal trends in the ISCA community. In Figure~\ref{fig:topichistory}, we see three types of long-term patterns. The first type of topics, such as \emph{data flow}, \emph{RISC}, \emph{parallel processing},\emph{instruction-level parallelsim}, \emph{multicore processor}, \emph{microprogramming}, \emph{branch prediction}, \emph{shared memory}, receive intensive coverage for a limited period of time. We suspect the interests in these topics subside as the topics are considered either mature or no longer of interest. Please note that the scale of the y-axis (the number of paper  abstracts mentioning the topic) varies among topics. The second type of topics, such as \emph{memory access}, \emph{speculate}, \emph{operating systems}, \emph{programming languages}, \emph{cache coherence}, \emph{CPU}, and \emph{memory systems}, receives periodic surge of coverage. Because these topics are about architecture support for key parts of computing systems, they drew research interests when the industry migrates to a new technology or new paradigm. The third type, such as \emph{GPU}, \emph{power consumption}, \emph{FPGA}, \emph{energy efficiency}, \emph{DRAM}, has received increasing coverage in recent years. Although they may belong to the first type when we look back a few years from now, it is too early to tell.


\section{Detailed Analysis of Trends Over the Years}
Each figure in this section represents a cloud of representative phrases that appeared in abstracts for a span of five years, except that the first and last figures cover a span of 3 years. We count the number of papers in which we find a match to our query phrase in the abstracts. We expect to have some small but potentially significant errors in the calculated weight for the phrases due to using only abstracts and some missing semantic relations between phrases in our natural language processing pipeline.
\subsection{ISCA in the 1970s}
\begin{figure}[h!]
    \begin{subfigure}[b]{\textwidth}
         \centering
         \includegraphics[width=\textwidth]{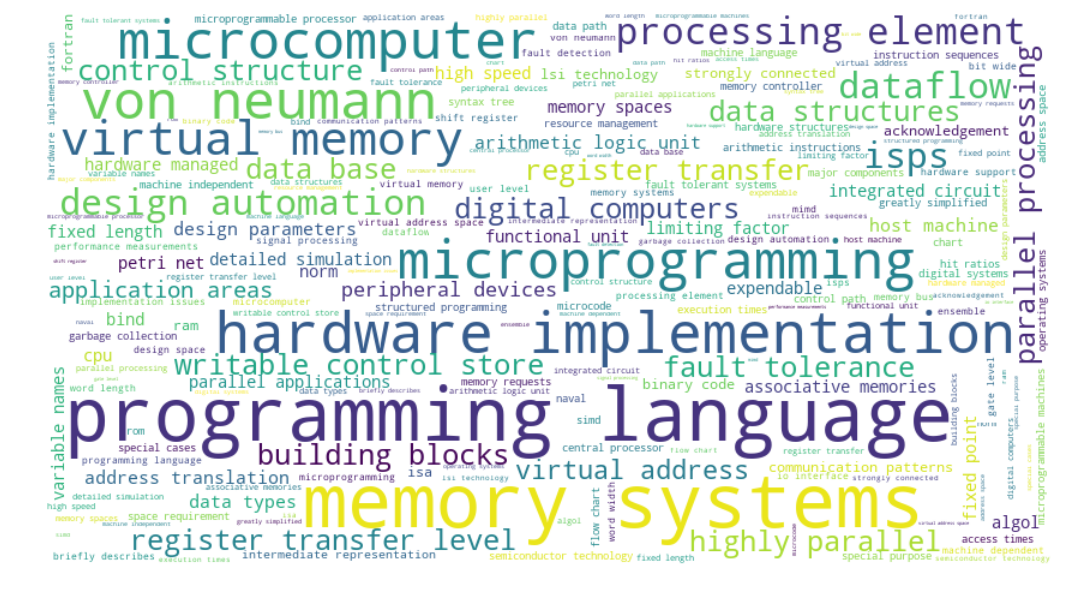}
         \caption{Word cloud visualization of topics - 1973-75}
         \label{fig:word-cloud-1973-1975}
     \end{subfigure}
     \hfill
     \centering
     \begin{subfigure}[b]{0.42\textwidth}
         \centering
         \includegraphics[width=\textwidth]{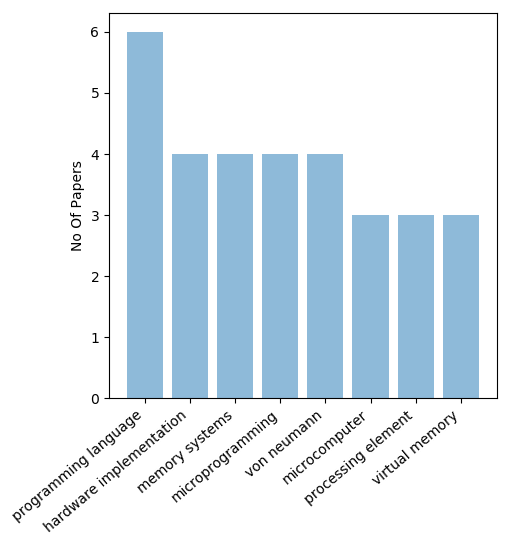}
         \caption{Top topics 1973-75}
        %  \label{fig:y equals x}
     \end{subfigure}
     \hfill
     \begin{subfigure}[b]{0.4\textwidth}
         \centering
         \includegraphics[width=\textwidth]{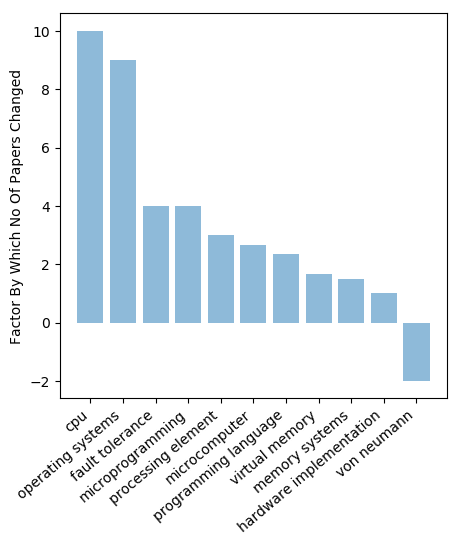}
         \caption{Topics with the most change in coverage from 1973-75 to 1976-80}
         \label{fig:delta1973-1975}
     \end{subfigure}
    \caption{Phrases and trends, 1973-1975.}
    \label{fig:1973-1975}
\end{figure}

The 1960s were the formative years of the computer industry, when only a few companies such as Burroughs, IBM, Control Data, UNIVAC, Wang Laboratories and Digital Equipment Corporation were able to produce products that were programmed in machine language and/or early high-level languages such as FORTRAN and COBOL. Later in 1970s, we see more industry players joining them such as CRAY Research. During this time,  many of the languages that we use today were developed. While high-level programming languages and compilers were being developed to improve the productivity, earlier efforts to map high-level languages to stored-program machines, commonly referred as the von Neumann architecture, resulted in high memory usage and long instruction sequences. Researchers began to propose designs where the programming languages are supported with hardware features. As shown in Figure~\ref{fig:1973-1975}, the phrases \emph{programming language} and \emph{hardware implementation}  were the No. 1 and No. 2 phrases during the period from 1973 to 1975.
The concept of Instruction Set Architecture (ISA) was pioneered by IBM in the 1960s. The same ISA that is implemented across multiple hardware generations allows the same software to work across different machine generations.  Binary codes from earlier IBM models such as 1401, 7040 and 7094 were also able to run unchanged on System/360 series.
%, which helped IBM to keep its customer base. 
During that time, microprogramming was the primary vehicle for a processor to interpret the instructions of the ISA and execute them on the native hardware. The success of the IBM System/360 product line solidified the role of ISA and microprogramming in the computer architecture community. 
%However, intermediate ISA code reduced the performance of native ISA. 
%When the control memory technology inside the processor became feasible IBM invested a lot in the technology with the primary goal to produce a range of processors with one ISA. 

%As a result, the architecture research community began to propose computer architecture and hardware implementations that can run programs in those high-level languages with improved efficiency. 
In the early 1970's, numerous designs were proposed to improve efficiency of running programs written
in those high-level languages, but only a handful were actually implemented \cite{hllc1}. Burroughs E-mode machines for Algol 60, Burroughs B2000 for COBOL, LISP machines and Intel 432 for Ada are some of those examples. A number of ideas based on microprogramming can be found in the ISCA papers published in those years that underline these efforts, which is reflected as the No. 2 ranking of the phrase \emph{microprogramming}. By 1969, the debate over virtual memory also concluded when IBM showed a consistent performance of their virtual memory overlay system over manual system. This was
also reflected in the ISCA publication during that time, where virtual memory was one of the top trends as shown in Figure~\ref{fig:1973-1975}.
% Thus, as shown in Figure~\ref{fig:1973-1975}, topics such programming languages, von Neumann, hardware implementation, memory systems, garbage collection, microcode generator, and register-transfer are prominently present during this time.

%Rising software development cost and falling cost of hardware with more transistors available on the circuit board made it attractive to some parts of architecture community to use newly available hardware resources for implementing part of the software. However, mapping more of software to hardware increased the hardware design and verification cost. For example, it took several years to debug SYMBOL system's hardware implemented language and operating system. As a result, we also see active coverage of topics such as operating systems, design automation, and fault tolerance.

\begin{figure}
    \begin{subfigure}[b]{\textwidth}
         \centering
         \includegraphics[width=\textwidth]{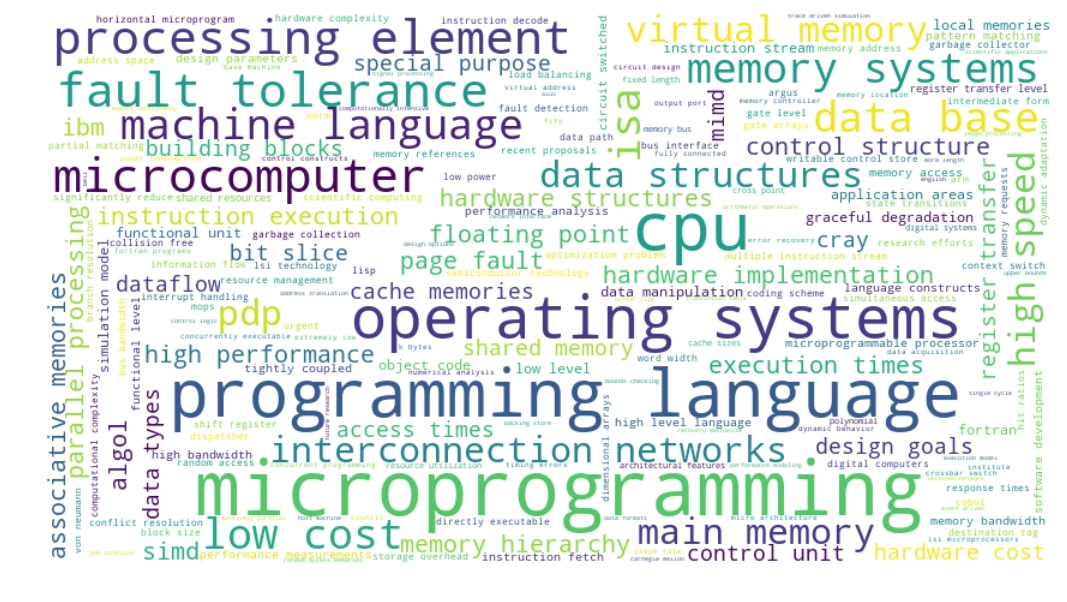}
         \caption{Word cloud visualization of topics - 1976-80}
        %  \label{fig:y equals x}
     \end{subfigure}
     \hfill
     \centering
     \begin{subfigure}[b]{0.42\textwidth}
         \centering
         \includegraphics[width=\textwidth]{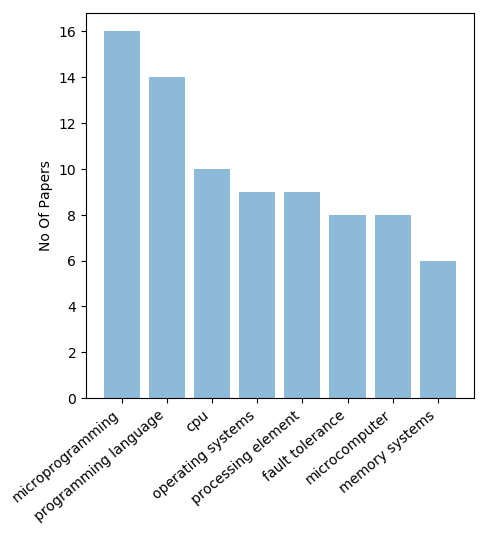}
         \caption{Top topics 1976-80}
        %  \label{fig:y equals x}
     \end{subfigure}
     \hfill
     \begin{subfigure}[b]{0.42\textwidth}
         \centering
         \includegraphics[width=\textwidth]{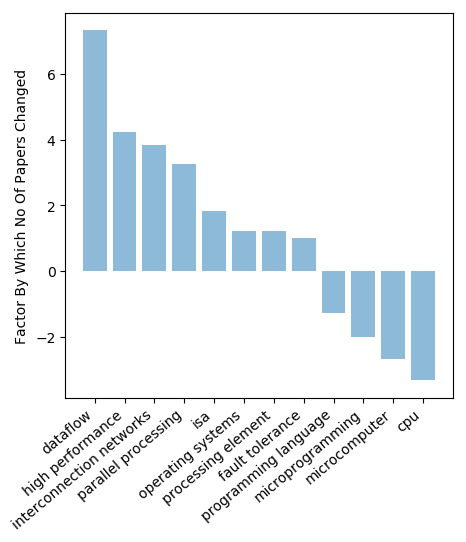}
         \caption{Topics with the most changes in coverage from 1976-80 to 1981-85}
        %  \label{fig:three sin x}
     \end{subfigure}
    \caption{Top Phrases and Trends, 1976-80}
    \label{fig:1976-1980}
\end{figure}

As more industry players design computer products in the 1970s, there was also an increased interest in developing new operating systems and providing improved support for operating systems that offer reliable, secure, and usable environments for the users. As a result, as shown in Figure~\ref{fig:delta1973-1975} for the late 1970s, the ISCA topics in operating systems, virtual memory and memory systems has also increased. Operating systems  continues to be one of the top trend in late 1970s as shown in Figure~\ref{fig:1976-1980}. 
There was also a burgeoning interest in parallel computing with topics like data flow, parallel processing, and processing elements. Both types of topics would increase in later years.

During 1970s, the computer industry went through a decade of innovation in the mini-computer movement. Companies such as Digital Equipment Corporation introduced mini-computers that are affordable by smaller companies and academic departments. Researchers and engineers can access these minicomputers through Cathode-Ray Tube (CRT) terminals on their desks rather than the card punchers in computing centers, which greatly improved their productivity. These minicomputers have new instruction sets that were implemented with microcode, which further stimulated the coverage of topics like CPU, instruction set, machine language, instruction execution, low cost, microprogramming, and writeable control.

With more accessibility to researchers, these mini-computers also accelerated the development of high-level languages. With the poor performance of high-level language implementations, one of the lessons learned was that it was not only about the language but also about computation, algorithm and memory access. Researchers began to investigate how these facets of a program can also be reflected in the ISA, which was implemented by a native hardware microarchitecture through microprogramming. Additionally, the desire to better support operating systems further motivated the introduction of sophisticated instructions that help data movement, security, and reliability. Previously, the control unit in a processor was hardwired as combinational logic and finite state machine. However, the need for more complex and powerful instructions made it difficult, time consuming and costly to design such hardwired processors. And since there were also very few CAD tools for hardware development and verification, this path was less productive in the late 1970s. This, in part, further contributed to the popularity of microprogramming at that time. Microcode simplified the processor design by allowing the implementation of control logic as a microcode routine stored in the memory instead of a dedicated circuit. The VAX ISA by Digital Equipment Corporation consisted of more than 300 instructions in its ISA. A number of complex instructions were introduced to support high-level languages and operating systems to bridge the semantic gap. However, it was observed that compilers rarely used those instructions and these instructions were kept just to support ``legacy codes'' in low-level libraries. The VAX polynomial-evaluate and CALL instructions are examples of such instructions. As shown in Figure~\ref{fig:1976-1980}, the period between 1976 and 1980 was the time when CPU design and microprogramming were at their peak. Programming languages, operating systems, processing element, and fault tolerance continue to receive strong attention.

\subsection{ISCA in the 1980s}
\begin{figure}[h!]
     \begin{subfigure}[b]{\textwidth}
         \centering
         \includegraphics[width=\textwidth]{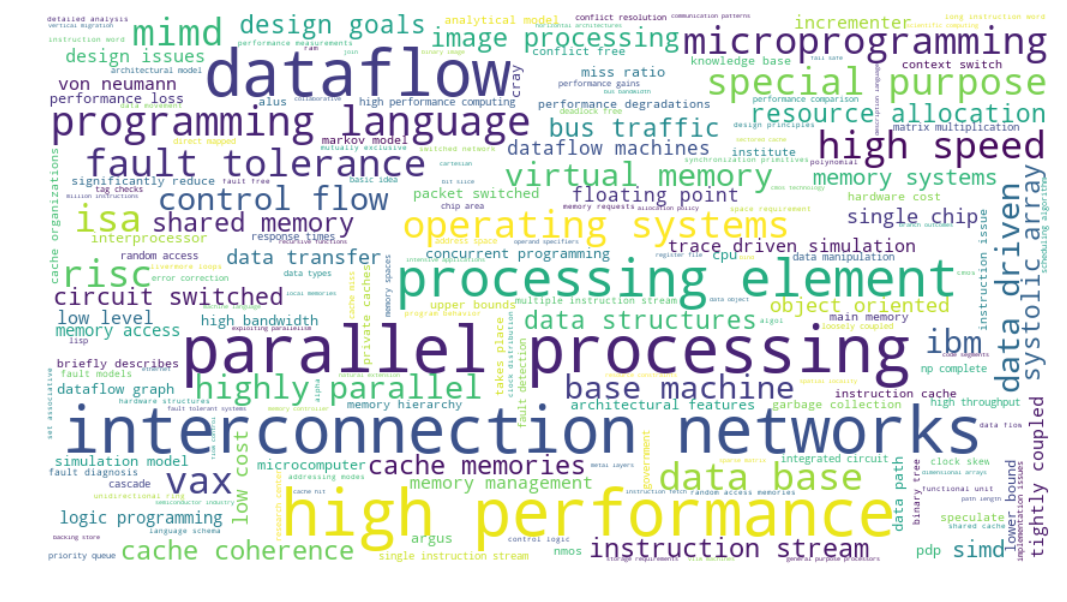}
         \caption{Word cloud visualization of topics - 1981-85}
        %  \label{fig:y equals x}
     \end{subfigure}
     \hfill
     \centering
     \begin{subfigure}[b]{0.42\textwidth}
         \centering
         \includegraphics[width=\textwidth]{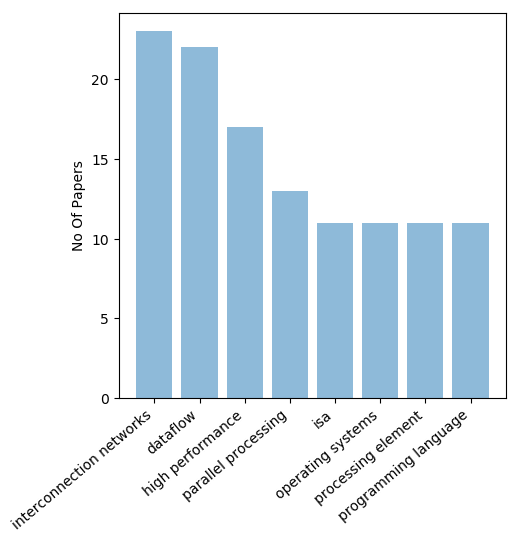}
         \caption{Top topics 1981-85}
        %  \label{fig:y equals x}
     \end{subfigure}
     \hfill
     \begin{subfigure}[b]{0.42\textwidth}
         \centering
         \includegraphics[width=\textwidth]{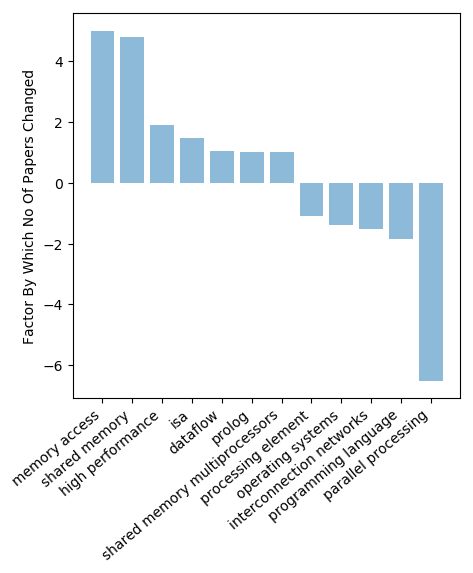}
         \caption{Topics with the most change in coverage from 1981-85 to 1986-90}
        %  \label{fig:three sin x}
     \end{subfigure}
    \caption{Top topics and trends, 1981-85}
    \label{fig:1981-1985}
\end{figure}

In the 1980s, the computer architecture community started to embrace parallel processing and high-performance computing. As shown in Figure~\ref{fig:1981-1985}, a great deal of attention by the ISCA community was paid to the interconnection network between processing elements.  The driving applications behind those systems were the  military and scientific applications including image processing, astrophysics and weather prediction. Individual processors were not capable of providing the required computation speed at that time. 
%In academia, Caltech Concurrent Computation project \cite{CCCP} built a MIMD supercomputer for scientific applications with 64 off-the-shelf processors, Intel 8086/8087 processors. (WMH: I removed this becasue it was strange that we singled this out and it is not even an ISCA paper.)

In industry, mini-supercomputers from Seqent and Alliant began to gain popularity. There was also a burgeoning interest in massively parallel systems such as Connection Machines from Kendal Square Research.  Three major components of those multiprocessing systems were processing elements, shared memory and interconnection network. In order to make efficient use of multiprocessing systems, it required ensuring a communication network that does not become the bottleneck. A number of ideas were introduced in ISCA publications around circuit switched networks, packet switched networks, multi stage networks, binary tree networks, bus traffic, resource scheduling and routing protocols. This is reflected in the No. 1 and No. 4 rankings of the \emph{interconnect network}
 and \emph{parallel processing} topics during early 1980s as shown in Figure~\ref{fig:1981-1985}. During this period, challenges in parallel programming have also motivated research in data flow architectures and data-driven execution where researchers proposed hardware mechanisms to identify instructions that are ready for execution and are scheduled for execution as soon as possible. This is reflected in the No. 2 ranking of the \emph{data flow} topic during this period. 

\begin{figure}
     \begin{subfigure}[b]{\textwidth}
         \centering
         \includegraphics[width=\textwidth]{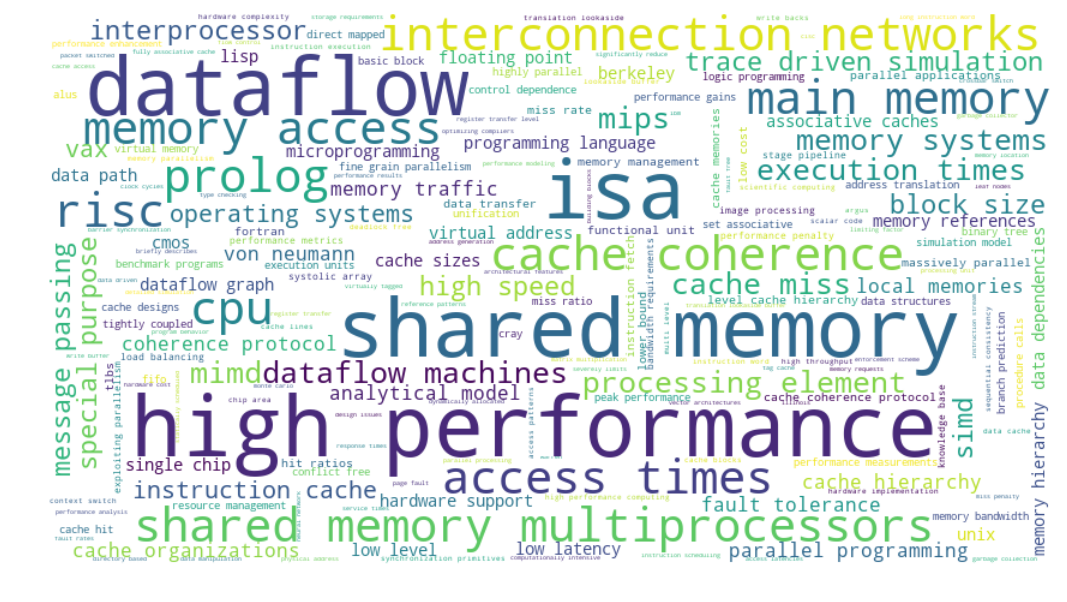}
         \caption{Word cloud visualization of topics from 1986-90}
     \end{subfigure}
     \hfill
     \centering
     \begin{subfigure}[b]{0.42\textwidth}
         \centering
         \includegraphics[width=\textwidth]{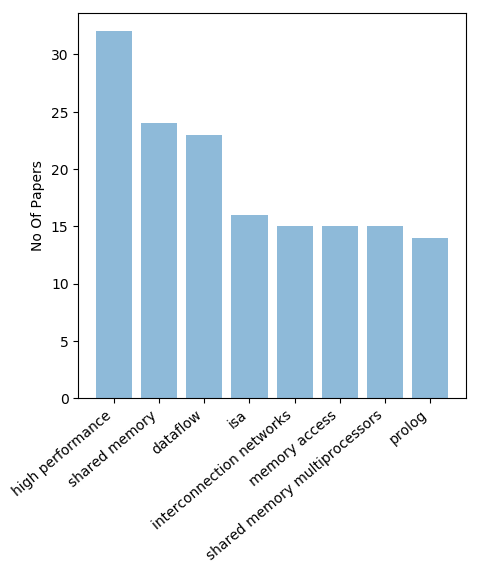}
         \caption{Top topics 1986-90}
         \label{fig:1986-1990(top)}
     \end{subfigure}
     \hfill
     \begin{subfigure}[b]{0.42\textwidth}
         \centering
         \includegraphics[width=\textwidth]{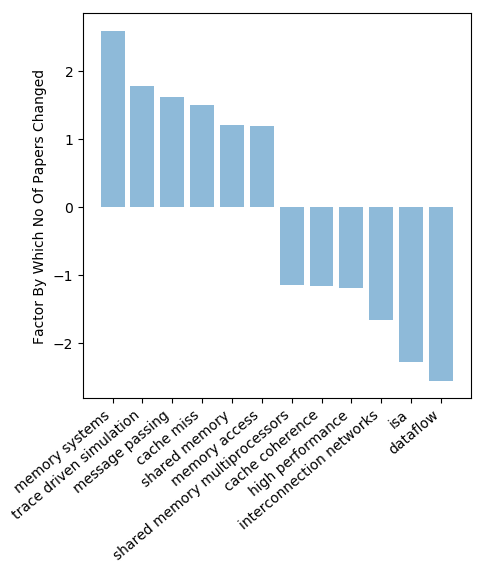}
         \caption{Topics with most changes in coverage from 1986-90 to 1991-95}
        %  \label{fig:three sin x}
     \end{subfigure}
    \caption{Top topics and trends, 1986-1990}
    \label{fig:1986-1990}
\end{figure}

Supporting shared memory in a parallel processing system has its own challenges, including system scalability to thousands of processors without dramatically increasing the memory access latency and memory interference. Levels of memory hierarchies were proposed to include caches for perfecting and reusing the data, which further required  developing coherence protocols and consistency models to reduce the burden on programmers in managing data values across the system. This is reflected in the significant increase in related topic coverage from 1981-85 to 1986-90, as shown in Figure~\ref{fig:1981-1985}(c). We also observe that, as shown in Figure~\ref{fig:1986-1990} and Figure~\ref{fig:1991-1995}, this trend continued even into the early 1990s. 

In the 1980s, the RISC movement, started with the CRAY-1 machine and the IBM 801 microprocessor project, advocated ISAs with simple instructions, which resulted in uniform instructions that are easier to decode and pipeline. The debate about the advantages and disadvantages of simpler instructions made \emph{ISA} one of the top topics during the period of 1986-90, as shown in Figure~\ref{fig:1986-1990}. 

% From 1986 to 1990, the industry began to produce microprocessor chips that contain most major components of traditional board-level CPUs. 
The RISC designs matched well with the transistor budget of the microprocessor chips during this time. Industry companies began to produce chips based on new ISAs like SPARC by SUN Microsystems, MIPS by MIPS, Inc., Spectrum by Hewlett-Packard, and 960 by Intel. However, the simpler instruction sets also increased the pressure for increased memory bandwidth for instruction fetch. Pipelining allowed the CPU clock frequency to improve much faster than that of the memory system. As a result, the memory system began to become an important bottleneck in overall system performance.  As a result, many researchers began to pay attention to memory accesses/references, virtual addresses, memory systems, main memory, and memory hierarchies. 

As the number of transistors further increased over time following the Moore's Law, there was also a resurgence of interest in the cache design. Although caches were used extensively in mainframe computers and minicomputers in the 1970s, the microprocessors started to have barely sufficient number of transistors in the 1980s to incorporate caches on chip to mitigate the memory access bottleneck. This was reflected in the increased ISCA coverage of topics such as cache memories, instruction caches, cache hierarchy, block sizes, cache misses, and trace-driven simulation for cache performance studies\ref{fig:cloud91-95}.

From 1986 to 1990, researchers also published intensively on architectures that supported Prolog, a programming language for rule-based inferences in artificial intelligence  applications, as shown in Figure~\ref{fig:1986-1990(top)}. This was mostly stimulated by the increased DARPA funding in the early 1980s for AI architecture projects in response to the Japanese 5th generation AI computers project. However, the interest in AI and Prolog would soon diminish due to the lack of practical AI applications.

\subsection{ISCA in the 1990s}
\begin{figure}
     \begin{subfigure}[b]{\textwidth}
         \centering
         \includegraphics[width=\textwidth]{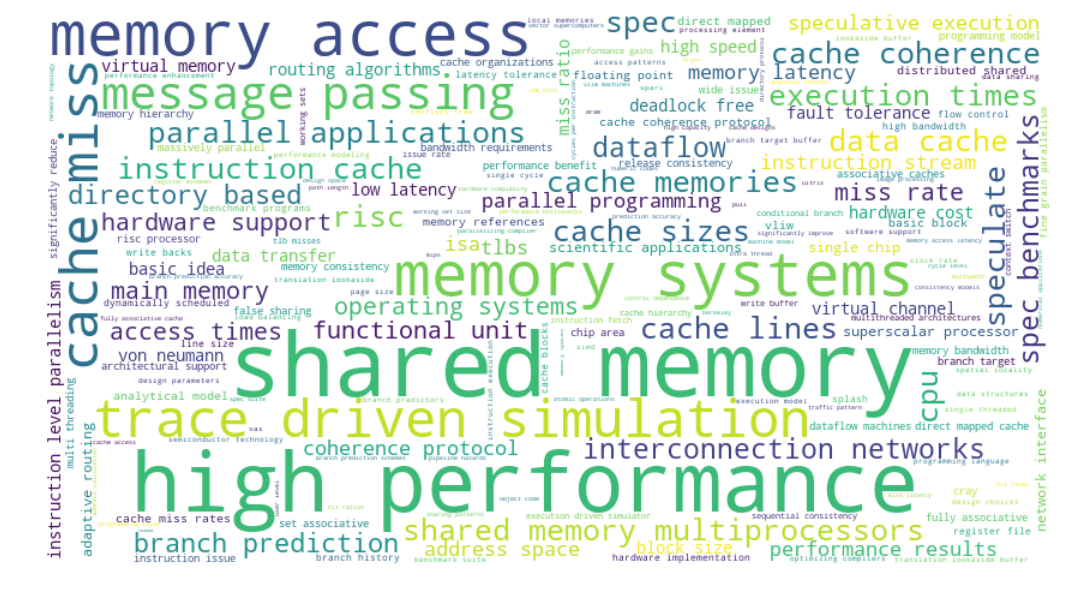}
         \caption{Word cloud visualization of topics - 1991-95}
         \label{fig:cloud91-95}
     \end{subfigure}
     \hfill
     \centering
     \begin{subfigure}[b]{0.4\textwidth}
         \centering
         \includegraphics[width=\textwidth]{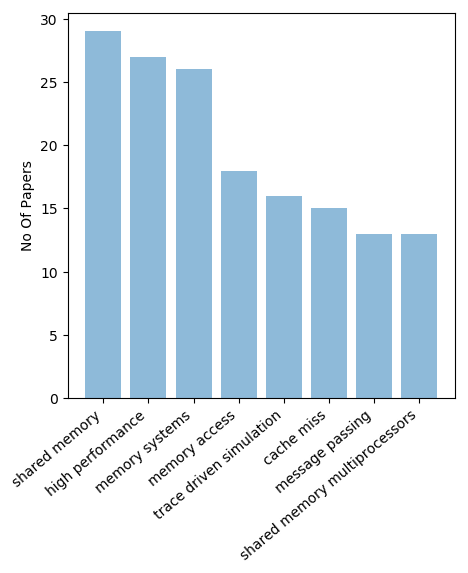}
         \caption{Top Topics 1991-1995}
        %  \label{fig:top91-95}
     \end{subfigure}
     \hfill
     \begin{subfigure}[b]{0.46\textwidth}
         \centering
         \includegraphics[width=\textwidth]{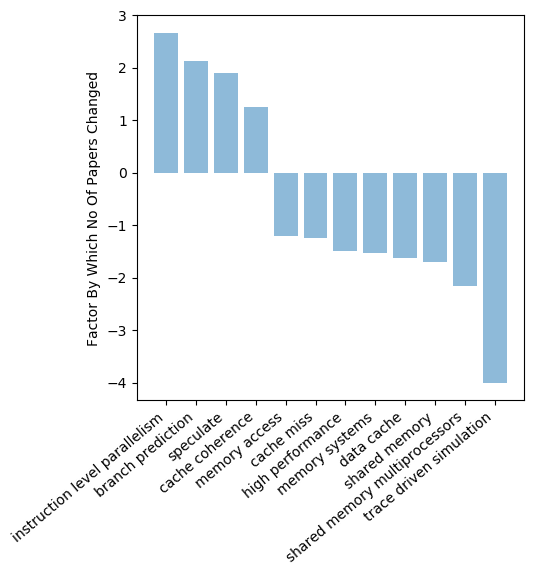}
         \caption{Topics with the most change in coverage from 1991-95 to 1996-2000}
        %  \label{fig:three sin x}
     \end{subfigure}
    \caption{Top topics and trends, 1991-95}
    \label{fig:1991-1995}
\end{figure}

In the early 1990s, the computer architecture researchers continue to publish extensively on \emph{shared memory} and \emph{shared memory multiprocessors}. The increasing commercial use of \emph{database} applications and scientific applications in the early 1990s further stimulated studies on \emph{shared memory} server designs and \emph{message-passing} clusters. Both trends eventually disrupted mainframes in the database market and the traditional vector supercomputers in the scientific computing market. However, it is also interesting to note that the coverage of these topics in ISCA will decrease dramatically in the next 5-year period.

\begin{figure}
     \begin{subfigure}[b]{\textwidth}
         \centering
         \includegraphics[width=\textwidth]{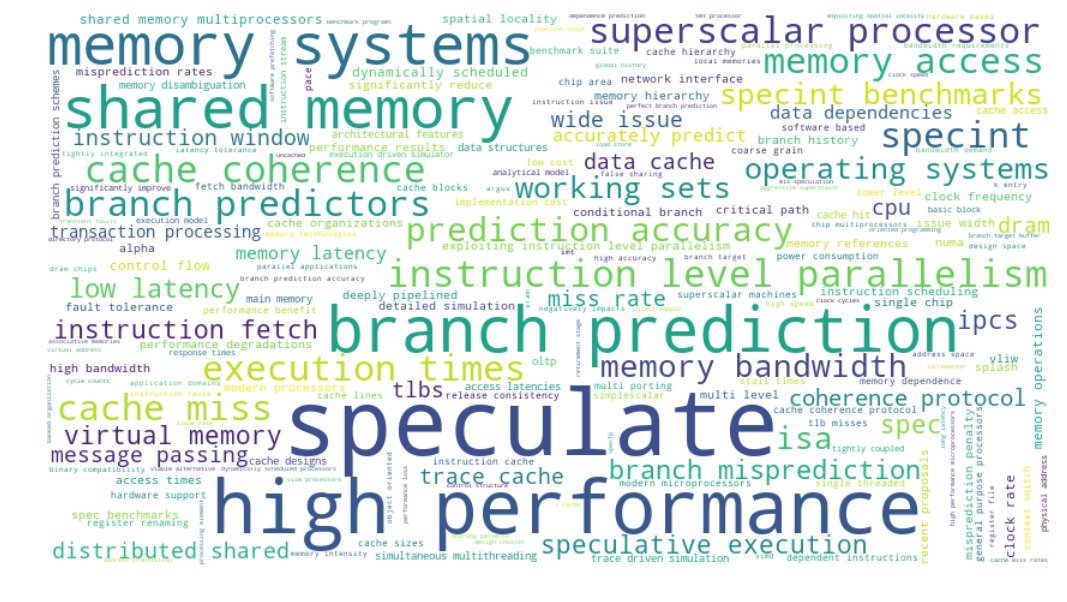}
         \caption{Word cloud visualization of topics - 1996-2000}
     \end{subfigure}
     \hfill
     \centering
     \begin{subfigure}[b]{0.42\textwidth}
         \centering
         \includegraphics[width=\textwidth]{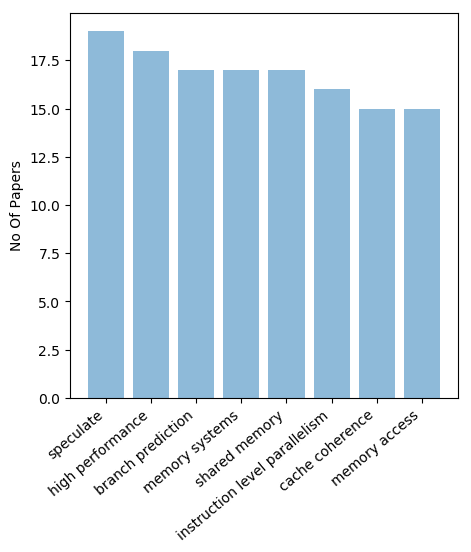}
         \caption{Top topics 1996-2000}
        %  \label{fig:y equals x}
     \end{subfigure}
     \hfill
     \begin{subfigure}[b]{0.42\textwidth}
         \centering
         \includegraphics[width=\textwidth]{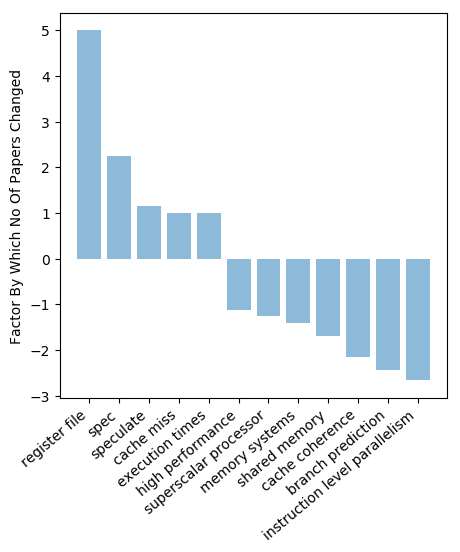}
         \caption{Topics with the most change in coverage from 1996-2000 to 2001-05}
        %  \label{fig:three sin x}
     \end{subfigure}
    \caption{Top topics and trends, 1996-2000}
    \label{fig:1996-2000}
\end{figure}

Thanks to the relentless scaling according to the Moore's Law, the the number of transistors available to the industry design teams increased to the level that allowed the designers to adopt techniques that had been only used in mainframe computers and supercomputers. Research in the 1980s set the foundation for exception handling in processors using out of order and speculative execution techniques.  During the 1990-95 period, as shown in Figure~\ref{fig:1991-1995}, computer architects used newly available transistors in building \emph{high-performance processors} with hardware schedulers for detecting \emph{instruction-level parallelism}, performing \emph{instruction re-ordering}, \emph{branch predictions} and \emph{speculative execution} together with \emph{pipelining} for increased \emph{clock frequency} and also to bridge the gap between memory latency and processing time. These innovations resulted in deeper pipelines and wider issue width in a generation of superscalar processors. 

The industry design of superscalar processors reached its peak during the late 1990s as Intel, AMD, MIPS/SGI, SUN, IBM, Hewlett-Packard all came up with superscalar microprocessor products in the mid 1990s.  As shown in Figure~\ref{fig:1996-2000}, \emph{high-performance} processor design techniques such as \emph{hardware speculation}, and \emph{branch prediction} received significant attention from the ISCA community during 1996-2000. The superscalar processors introduced at that time included Intel Pentium, MIPS R8000 and IBM Power series.

However, the increased clock frequency and execution throughput of these processors placed even more pressure on the \emph{memory system}, which motivated more studies on \emph{memory access}, \emph{data caches}, \emph{cache sizes}, \emph{block sizes}, and \emph{miss rates/ratios}. During this time, the computer architecture community started to converge on using trace driven simulation based on \emph{SPEC benchmarks} in studying processor pipelines as well as \emph{cache memories}. 

It is interesting to note though that, as shown in Figure~\ref{fig:1996-2000}(c), the coverage of \emph{instruction-level parallelism}, \emph{branch prediction}, \emph{superscalar processor} \emph{shared memory}, and \emph{memory systems} would drop significantly in the next 5-year period.

% \begin{figure}[h!]
%   \centering
%     \includegraphics[width=\linewidth]{figures/1995.png}
%     \caption{1996-2000}
%   \label{fig:coffee}
% \end{figure}

%Not only in the context of single processor system, speculative execution was also studied in the context of shared memory multiprocessing systems in ISCA. 

The decade from 1991 to 2000 was a dark age for research in massively parallel computing systems and special-purpose acceleration hardware. The industry not only rode on the Moore's Law with the exponentially increasing number of transistors, but also started to deviate from the Dennard's scaling, trading more power consumption for super-linear performance improvement over time. This strategy resulted in such fast advancement in microprocessor performance that it eclipsed any benefit of massively parallel computing systems or special-purpose hardware accelerators. Thus, the term "Killer Micros" became prominent, indicating the fact that the fast advancing microprocessor performance killed off the research and development of massively parallel computing and special-purpose hardware acceleration. But all of this would change in the next decade as we'll discuss below.

\subsection{ISCA in the 2000s}
% \begin{figure}[h!]
%   \centering
%     \includegraphics[width=\linewidth]{figures/2000.png}
%     \caption{2001-2005}
%   \label{fig:coffee}
% \end{figure}
\begin{figure}
     \begin{subfigure}[b]{\textwidth}
         \centering
         \includegraphics[width=\textwidth]{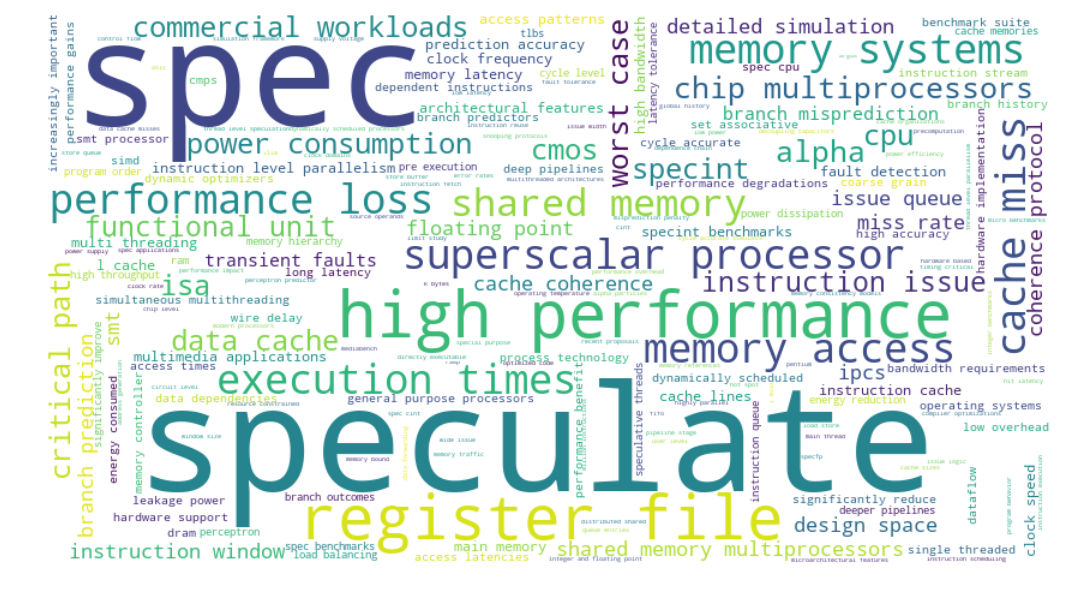}
         \caption{Word cloud visualization of topics 2001-05}
     \end{subfigure}
     \hfill
     \centering
     \begin{subfigure}[b]{0.4\textwidth}
         \centering
         \includegraphics[width=\textwidth]{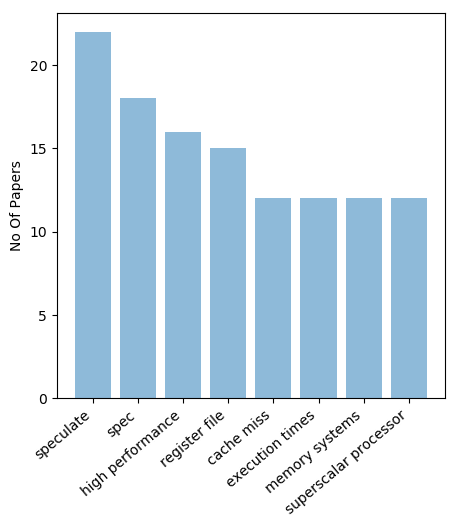}
         \caption{Top topics 2001-05}
        %  \label{fig:y equals x}
     \end{subfigure}
     \hfill
     \begin{subfigure}[b]{0.45\textwidth}
         \centering
         \includegraphics[width=\textwidth]{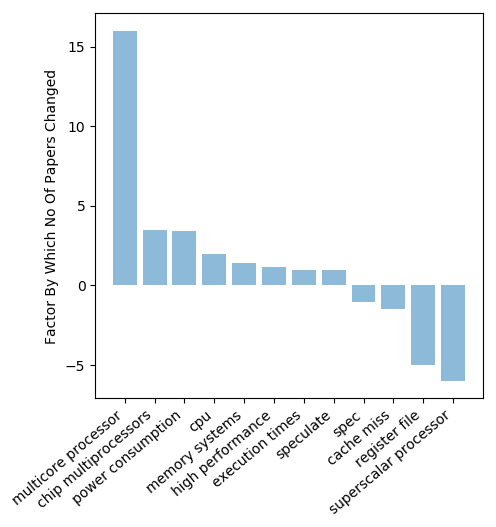}
         \caption{Topics with the most change in coverage from 2001-05 to 2006-10}
        %  \label{fig:three sin x}
     \end{subfigure}
    \caption{Top topics and trends, 2001-2005}
    \label{fig:2001-2005}
\end{figure}

The fact that \emph{SPEC} became the top topic phrase for the period of 2001-05 indicated that by this time the computer architecture community has fully embraced quantitative approaches to computer architecture research. We observe that \emph{SPEC CPU2000}, or its predecessor SPEC CPU95, has became the de facto standard  for measuring processor and/or memory-hierarchy performance in 2000s. The benchmarks were used to measure a wide variety of system designs, including \emph{multiprocessing systems} with a multi-level \emph{memory hierarchy}, \emph{memory address translation} overheads in cloud and server applications, a period when internet was coincidentally getting popular. 

The period of early 2000s also saw the peak of \emph{speculation techniques} in superscalar processors, VLIW processors, and memory systems. The industry was producing VLIW/EPIC processors such as the Intel Itanium. Computer architecture researchers were publishing extensively on architectural support for compile-time control \emph{speculation} and data \emph{speculation}. Researchers also published extensively on \emph{register file} design for both VLIW/EPIC and wide-issue superscalar processors. These processors require very large number ports to support simultaneous accesses to the register file by many instructions at different stages of the processor pipeline. This triggered the coverage of large register files with multiple read and write ports. A number of ISCA publications at that time looked at  various aspects of register file design, including their organization, access time, power consumption and cost.  

From 1995 to 2005, the industry has been achieving super-linear  scaling of \emph{high performance} designs, especially the clock frequency, over time at the cost of increasing power consumption. As we mentioned earlier, this is accomplished by deviating from the Dennard scaling principle of linearly scaling the performance in each generation of technology while keeping power consumption constant. By 2005, the power consumption of microprocessors has reached the limit of practical heat dissipation mechanisms.   As a result, computer architecture researchers began to focus on \emph{energy efficiency}, which would be one of the highly ranked ISCA topics with the most increased coverage from 2001-05 to 2006-10, as shown in Figure~\ref{fig:2001-2005}(c). On the other hand, the coverage of superscalar processors and register files would drop significantly in the next 5-year period. It is interesting to note that Figure~\ref{fig:2001-2005}(c) presents one of the most dramatic shift of topic coverage throughout the ISCA history.
%parallelism in a single instruction, \emph{instruction window} size, optimal \emph{pipeline depth}, \emph{DRAM system performance}, \emph{cache performance} and \emph{memory compression}. 
%Industry was trying to build faster and faster systems by already exploiting multiprocessing, pipelining and faster clock speed, remember superscalar processors. However, there also came some structural hazards such as the requirement for parallel execution units for pipelining and parallel access to registers and caches. 

% \begin{figure}[h!]
%   \centering
%     \includegraphics[width=\linewidth]{figures/2005.png}
%     \caption{2006-2010}
%   \label{fig:coffee}
% \end{figure}

\begin{figure}
     \begin{subfigure}[b]{\textwidth}
         \centering
         \includegraphics[width=\textwidth]{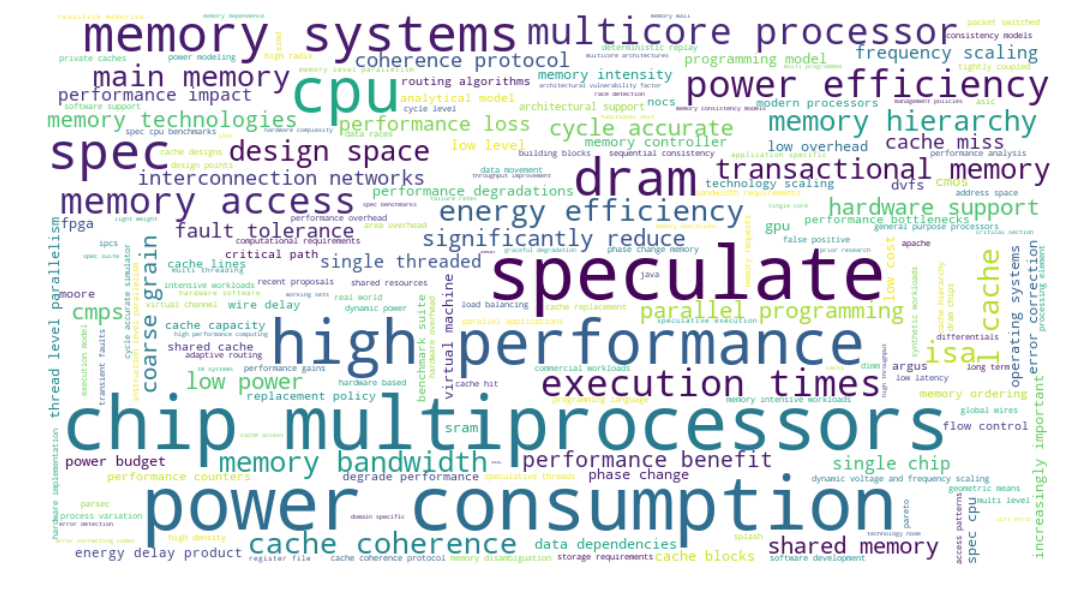}
         \caption{Word cloud visualization of topics - 2006-2010}
     \end{subfigure}
     \hfill
     \centering
     \begin{subfigure}[b]{0.42\textwidth}
         \centering
         \includegraphics[width=\textwidth]{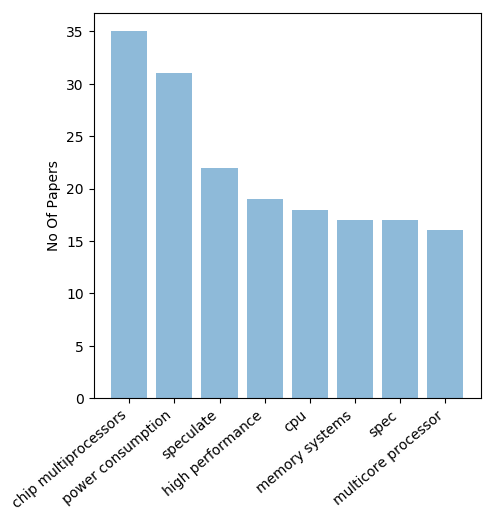}
         \caption{Top topics - 2006-2010}
        %  \label{fig:y equals x}
     \end{subfigure}
     \hfill
     \begin{subfigure}[b]{0.4\textwidth}
         \centering
         \includegraphics[width=\textwidth]{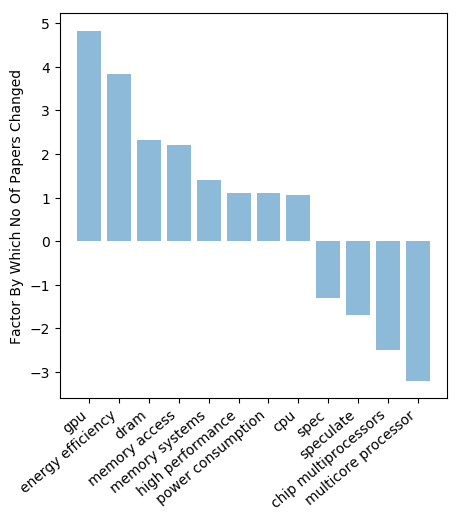}
         \caption{Topics with the most change in coverage from 2006-10 to 2011-15}
        %  \label{fig:three sin x}
     \end{subfigure}
    \caption{Top topics and trends, 2006-10}
    \label{fig:2006-2010}
\end{figure}

The period of 2006-2010 is the start of the era of \emph{chip multiprocessors}, a.k.a. \emph{multicore processors}. Before the availability of commercial multicore processors, superscalar processors were packed with more and more \emph{functional units}. The instructions were dispatched to the available functional units. However, this way of scaling the performance hit a practical barrier as the industry design teams struggled to exploit sufficient instruction-level parallelism to productively utilize additional functional units for increased performance. The industry made a major pivot from uni-processor clock frequency and instruction-level parallelism scaling to multicore scaling around 2003. The clock frequencies and instruction-level parallelism of each CPU core will largely remain the same, whereas the number of cores would increase over time. In fact, in some designs, the clock frequency may even be reduced to reduce power consumption to accommodate more cores for a given power budget. This turn away from clock frequency and instruction-level processing was reflected in the reduced coverage of topics like \emph{superscalar processor} and ~\emph{register file} from 2001-05 to 2006-10, as shown in Figure~\ref{fig:2001-2005}(c).

IBM launched Power4 that was the first dual core processor in 2001. Compaq developed piranha system for high-performance servers by integrating eight simple Alpha processor cores along with a two-level cache hierarchy onto a single chip. Sun Microsystems launched Niagra in 2005 that was eight-core Web server CPU. On-chip multiprocessing systems were thus studied in detail by the ISCA community in varying contexts, including cache hierarchies, power consumption, communication overheads, thread to core assignment for Simultaneous Multithreading (SMT), soft errors under low voltage, cache coherence protocols, interconnect networks, QoS, task scheduling and power management. The strong interests from the ISCA research community to support this movement were reflected in  the high coverage of topics such as \emph{chip multiprocessors}, \emph{multicore processors}, \emph{power consumption}, and \emph{energy efficiency} during the period of 2006-10, as shown in Figure~\ref{fig:2006-2010}. It is also interesting to note that the term \emph{chip multiprocessor} gave way to \emph{multicore processor}, which was reflected in the drop of their coverage from 2006-10 to 2011-15 as shown in Figure~\ref{fig:2006-2010}(c).

%Interestingly we can also see some interest in ISCA community in building machine learning based DRAM controllers from multi-core chips in 2008. Java runtime parallelizing machine (Jrpm), a complete system for parallelizing sequential programs automatically was introduced. Sun Microsystems proposed simultaneous speculative threading (SST) for its on-chip multicore ROCK processor.  

\subsection{ISCA in the 2010s}

% \begin{figure}[h!]
%   \centering
%     \includegraphics[width=\linewidth]{figures/2010.png}
%     \caption{2011-2015}
%   \label{fig:coffee}
% \end{figure}
\begin{figure}
    \begin{subfigure}[b]{\textwidth}
         \centering
         \includegraphics[width=\textwidth]{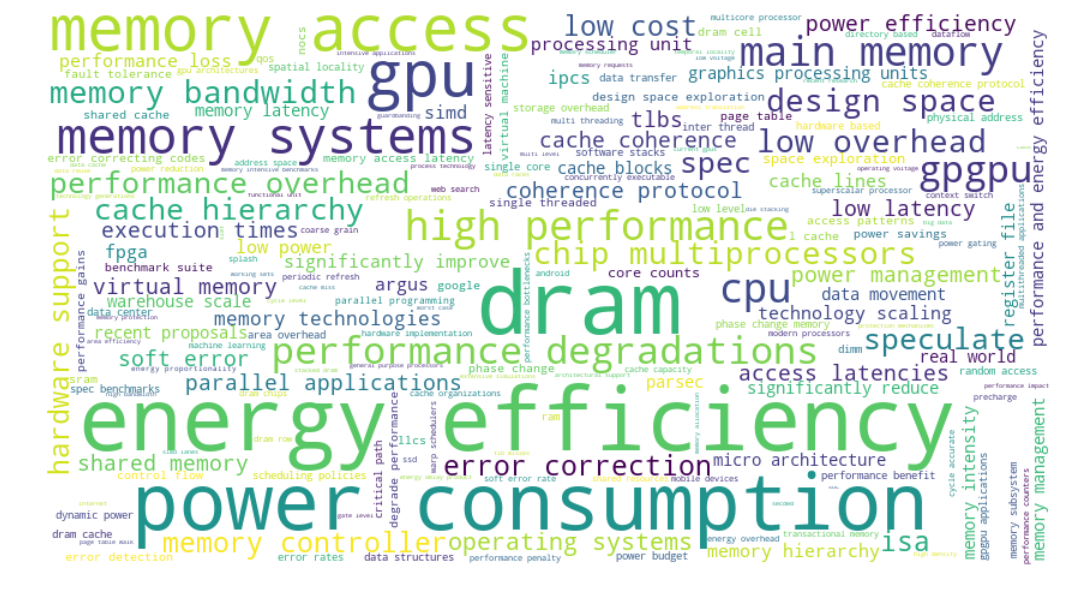}
         \caption{Word cloud visualization of topics, 2011-2015}
         \label{fig:2011-2015}
     \end{subfigure}
     \hfill
     \centering
     \begin{subfigure}[b]{0.4\textwidth}
         \centering
         \includegraphics[width=\textwidth]{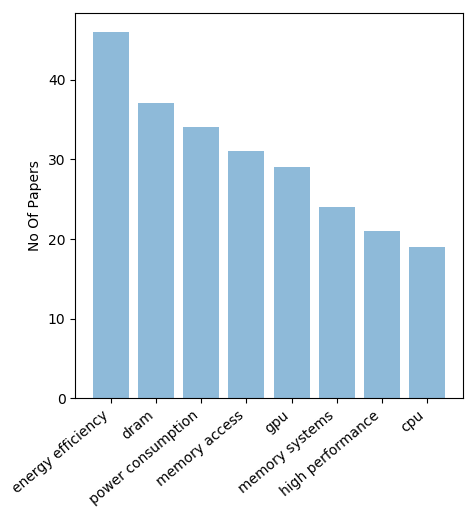}
         \caption{Top topics, 2011-2015}
        %  \label{fig:y equals x}
     \end{subfigure}
     \hfill
     \begin{subfigure}[b]{0.4\textwidth}
         \centering
         \includegraphics[width=\textwidth]{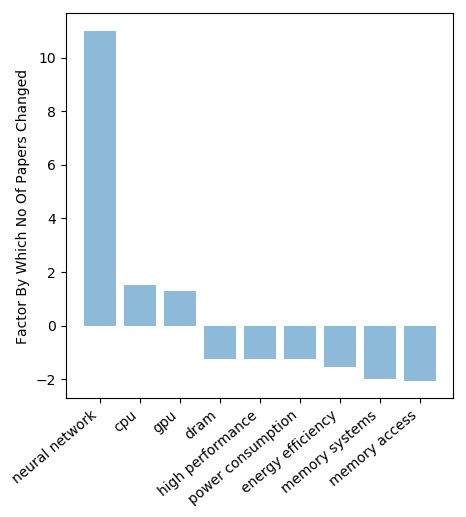}
         \caption{Topics with the most change in coverage from 2011-15 to 2016-18}
        %  \label{fig:three sin x}
     \end{subfigure}
    \caption{Top Topics and Trends 2011-2015}
    \label{fig:2011-2015}
\end{figure}

During the 2011-15 period, as shown in Figure~\ref{fig:2011-2015}, \emph{power consumption} and \emph{energy efficiency} became the top topics for ISCA authors.  Meanwhile, advances in the mobile devices and internet technology fueled an exponential growth in the tech industry with a variety of applications that generated huge amount of data. GPUs with hundreds of processing cores, already in use by the industry for graphics processing, proved to be high throughput devices for processing large amount of data. They became more general purpose with the introduction of CUDA programming model in 2007. The GPU equivalent of CPU cores, or Streaming Multiprocessors, run at about half of the clock frequency as CPU cores to achieve higher energy efficiency. The savings in the power consumption enabled the GPU designers to provision much higher memory bandwidth and thread-level parallelism. A major challenge was to program these massively parallel processors. 

A movement to empower  application developers to develop parallel applications started in 2007 with libraries and education materials from NVIDIA and academic institutions such as the University of Illinois at Urbana-Champaign, the University of California, Davis and the University of Tennessee, Knoxville. By 2011, there had been strong momentum in GPU libraries and applications in High-Performance computing. During the period of 2011-15, China, US, and Japan began to build top supercomputers based on CUDA GPUs. Examples were Tienhe 1 in China, Tsubame at Toyo Tech, Titan in Oakridge National Lab, and Blue Waters at the University of Illinois at Urbana-Champaign. 

These powerful GPU solutions and the CUDA programming model support also paved the way for machine learning. Using CUDA GPUs, a team from the University of Toronto  trained the AlexNet using 1.2 million images and won the ImageNet competition in 2012 with a large margin against the team in the 2nd place. This victory ignited wide interests in Neural Networks for computer vision and other cognitive applications. Nvidia used it to capitalize its investment and quickly developed cuDNN. Several other fields such as personalized medicine, genomics, physics, economics also realized that GPUs may help to consume lots of existing data for scientific breakthroughs. Some of the Top10 supercomputers also got equipped with GPUs. However, power consumption for large computing clusters was still a bottleneck. In 2011-15, we saw that power consumption has been the biggest concern, and
the number of publications in ISCA increased to address this challenge in the context of data centers. Research ideas and solutions were
proposed by the ISCA community from industry and academia at different layers of computing stack, circuits, architecture and algorithms.

% \begin{figure}[h!]
%   \centering
%     \includegraphics[width=\linewidth]{figures/2015.png}
%     \caption{2016-2018}
%   \label{fig:coffee}
% \end{figure}
\begin{figure}
    \begin{subfigure}[b]{\textwidth}
         \centering
         \includegraphics[width=\textwidth]{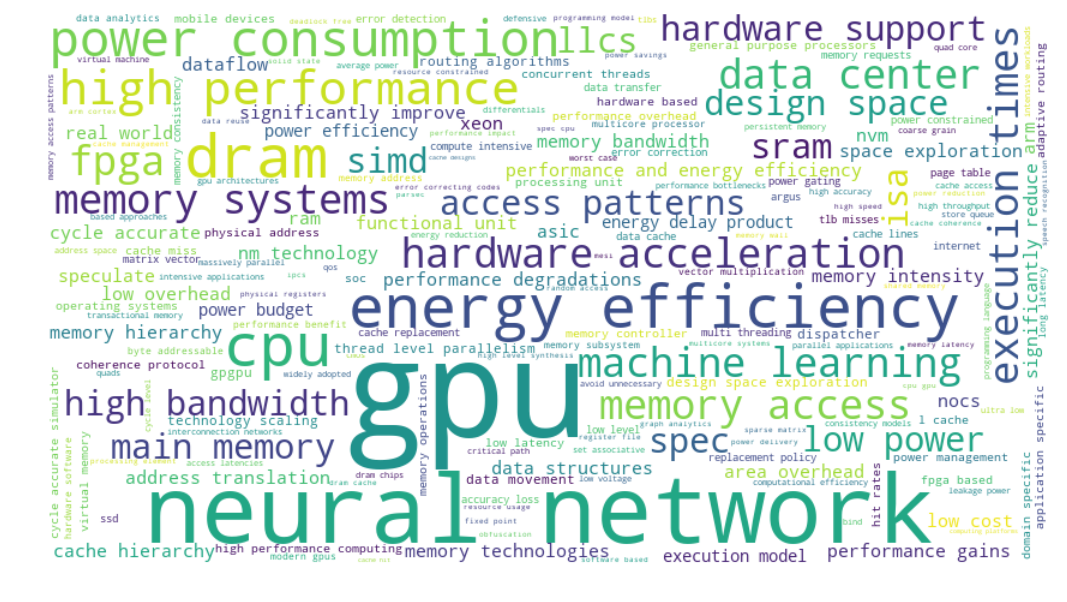}
         \caption{Visualization of Phrases from 2016-2018}
         \label{fig:wordcloud-2016-2018}
     \end{subfigure}
     \hfill
     \centering
     \begin{subfigure}[b]{0.4\textwidth}
         \centering
         \includegraphics[width=\textwidth]{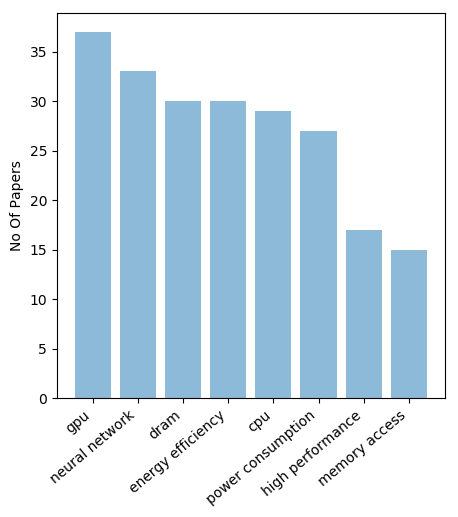}
         \caption{Top topics, 2016-2018}
        %  \label{fig:y equals x}
     \end{subfigure}
    %  \hfill
    %  \begin{subfigure}[b]{0.4\textwidth}
    %      \centering
    %      \includegraphics[width=\textwidth]{figures/TechtrendDelta2018.png}
    %      \caption{Delta:2011-2015 \& 2016-18}
    %     %  \label{fig:three sin x}
    %  \end{subfigure}
    \caption{Top topics and trends, 2016-18}
    \label{fig:2016-2018}
\end{figure}

In the most recent period of 2016-18, we saw that  machine learning based on Neural Networks have made their way to many applications and have become part of real-life systems. In Figure\ref{fig:wordcloud-2016-2018}, we see such influence in the ISCA publication trends. Computer architects acted quickly and addressed the related challenges to build machines that can process cognitive workflows in an efficient manner. Naturally, because of the increasing trend in machine learning and Neural Networks, GPUs have thus gained tremendous attention from the ISCA community. GPUs have made the training and learning more practical in terms of time and energy consumption. The desire to train more models has motivated the development of specialized hardware for tensor processing, referred to as TPUs and Tensor Cores.

However, processing large amount of data with GPUs and tensor processing hardware is efficient only when there is enough reuse of data primarily because of the memory wall. For applications with random data access pattern and poor cache hit rate, GPUs are not suitable. Even for applications with regular access patterns, what we have witnessed so far is under utilized GPUs because of not enough reuse in the applications. As data is growing exponentially, we should also expect an exponential growth in storage requirement, data movement and energy consumption. All these concerns are clearly visible in the ISCA publication trends during 2016-18. Performance from any of the existing technologies has not shown to be scalable to process this large amount of data within the required power and throughput budget. Radical changes are required from top to bottom both in software and hardware. 

\section{Outlook of Future Computer Architecture Research}
One of the key questions to the computer architecture community in the coming decade is how computer architects will address  challenges to scale the performance by 100x while staying in the required power and cost budgets? For storing the large amount of data to be processed,  we expect there may be a departure from existing storage devices to relatively lower cost and lower latency storage solutions. SSD is already used widely in consumer electronics and is now making its way into high performance computing solutions. Another trend is for main memory to be extended through 3D integration. Today top-of-the-line NVIDIA GPUs have already equipped with high bandwidth stacked DRAM solutions. In order to restrict the amount of data movement, an upcoming trend can be observed towards in-memory-computing and near-memory-computing solutions. 

In a few years, we expect to witness a compute hierarchy parallel to the existing memory hierarchy. However, a million-dollar question would be how to address the complexity of compute logic at different levels of the compute hierarchy given a variety of applications. One of the solutions along this line is to put logic layer underneath a stacked DRAM, which has captured a lot of attention from industry and academia. However, what compute logic goes where in the compute hierarchy is still an open research question. We will also likely witness parallelism by having a massive number of simple, energy efficient processing cores running at lower clock frequency, distributed not only across all the levels of memory hierarchy but at multiple levels in a memory die to reach bandwidth at the scale of TB/s. One fundamental idea is to restrict the data movement to local compute units if possible, increase parallelism, lower power consumption, and achieve high bandwidth  and high throughput. 

We will continue to see accelerators working along with general purpose processors. In the end, the whole system would consist of heterogeneous computing cores interconnected via an interconnection fabric. One of the daunting challenges is to program such devices, which requires  researchers rethinking of the whole software stack to ease  programmers' life. Since late 1970s, we have been developing high-level languages. We think that computer scientists are now going to have another iteration of the design cycle. We will see the development of even higher-level languages including  domain specific languages (DSLs) and new compilers that would support a variety of those domain specific languages and hardware platforms using a hierarchy of intermediate representations. In the next few years, we will also likely see an increasing interest in using machine learning as a tool to depart from the paradigm of hand crafted rules to semi-automation. It may help compilers, schedulers, branch predictors and any other computing system operations to learn from application behavior running in the cloud or at the edge. However, benefits of such a paradigm shift are still to be investigated.   

\section{About this Study}
This study began with a research project, called DISCVR, conducted at the IBM-ILLINOIS Center for Cognitive Computing Systems Reseach (c3sr.com). The goal of DISCVR was to build a practical NLP based AI solution pipeline to process large amounts of PDF documents and evaluate the end-to-end NLP capabilities in understanding large amount of unstructured data such as PDF files. This will in turn help us better understand the computation
patterns and requirements of modern AI solutions on the underlying computing systems.
While building such a prototye, an early use case came to us thanks to the
2017 IEEE/ACM International Symposium on Microarchitecture (MICRO-50) Program Co-chairs, Drs. Hillery Hunter and Jaime Moreno. They asked us if we can perform some data-driven analysis of the past 50 years MICRO papers and show
some interesting historical perspectives on MICRO's 50 years of publication. Because of the limited amount of time, we were only able to produce some
preliminary results and delivered an invited talk during that year's MICRO opening reception. It generated some interesting discussions, but the lack of insights from those early results has provided limited use of that work.
That undertake has, however, planted a seed in our C3SR center. We learned two important lessons from that experience: (1) building an AI solution to truly understand unstructured data is hard in spite of the many claimed successes in natural language understanding; and (2) providing
a data-driven perspective on computer architecture research is a very
interesting and fun project. 

Since then, we continued to push those two frontiers of research at the C3SR center. On the first topic, we built a prototype of paper review matching system, called C3SR System for Reviewer Assignment (CSRA), and used that
system to help the Program Co-chairs of 2019 ISCA, Drs. Hillery Hunter and Erik Altman, with their paper review assignment task. On the second topic,
we decided to conduct a more thorough study based on all past ISCA papers, which resulted this article.

We recognize that we have just scratched the surface of 
natural language understanding of unstructured data, and there are many more
aspects that we can improve (and we are still working on them). But even with
our current study, we felt there were enough interesting findings that may
be worthwhile to share with the community. Hence we decided to write this
article to summarize our findings so far based only on ISCA publications. Our hope is to generate
further interests from the community in this topic, and we welcome
collaboration from the community to deepen our 
understanding both of the computer architecture research 
and of the challenges of NLP-based AI solutions. For example, a similar study
can be conducted for other conferences (such as MICRO) and in other research areas (such as CVPR and SIGKDD).

\section{Acknowledgement}
We would like to thank Mr. Abdul Dakkak and Ms. Cheng Li from the C3SR center
for their early work on DISCVR and their contribution to the MICRO-50
data analysis. We would also like to thank Mr. Jong Yoon Lee, Ms. Hongyu Gong and Ms. Renee Zhu from the C3SR center for their contributions to the CSRA project, which
had direct impact on the data analysis as used in this article. We would also like to thank Drs. Hillery Hunter, Jaime Moreno and Erik Altman from IBM Research for their encouragement and feedback for our work.  

{}
\end{document}